\documentclass[twocolumn]{aastex63}

\newcommand{\Msun}{M_{\sun}}
\newcommand{\Mwd}{M_{\mathrm{WD}}}
\newcommand{\Rwd}{R_{\mathrm{WD}}}
\newcommand{\Mbh}{M_{\mathrm{BH}}}

\newcommand{\silicon}{^{28}\mathrm{Si}}

\newcommand{\calcium}{^{40}\mathrm{Ca}}
\newcommand{\nickel}{^{56}\mathrm{Ni}}

\newcommand{\kmpersec}{\mathrm{km}\,\mathrm{s}^{-1}}
\newcommand{\ergpersec}{\mathrm{erg}\,\mathrm{s}^{-1}}

\newcommand{\secref}[1]{Section~\ref{#1}}
\newcommand{\figref}[1]{Figure~\ref{#1}}

\received{September 20, 2019}
\revised{November 20, 2019}
\accepted{}
\submitjournal{ApJL}

\shorttitle{Helium WD TDE emission}
\shortauthors{Kawana et al.}

\begin{document}

\title{Rapid Transients Originating from Thermonuclear Explosions in Helium White Dwarf Tidal Disruption Events}

\correspondingauthor{Kojiro Kawana}
\email{kawana@utap.phys.s.u-tokyo.ac.jp}

\author[0000-0002-5942-7485]{Kojiro Kawana}
\affiliation{Department of Physics, School of Science, The University of Tokyo, 7-3-1, Bunkyo, Tokyo 113-0033, Japan}

\author[0000-0003-2611-7269]{Keiichi Maeda}
\affiliation{Department of Astronomy, Kyoto University, Kitashirakawa-Oiwake-cho, Sakyo-ku, Kyoto, 606-8502, Japan}

\author[0000-0001-7925-238X]{Naoki Yoshida}
\affiliation{Department of Physics, School of Science, The University of Tokyo, 7-3-1, Bunkyo, Tokyo 113-0033, Japan}
\affiliation{Kavli Institute for the Physics and Mathematics of the Universe (WPI), The University of Tokyo, 
5-1-5, Kashiwanoha, Kashiwa, Chiba 277-8583, Japan}
\affiliation{Research Center for the Early Universe, School of Science, The University of Tokyo, 7-3-1, Bunkyo, Tokyo 113-0033, Japan}

\author[0000-0002-8461-5517]{Ataru Tanikawa}
\affiliation{Department of Earth Science and Astronomy, College of Arts and Sciences, The University of Tokyo, 
3-8-1 Komaba, Meguro-ku, Tokyo 153-8902, Japan}
\affiliation{RIKEN Advanced Institute for Computational Science, 7-1-26 Minatojima-minami-machi, Chuo-ku, Kobe, Hyogo 650-0047, Japan}

\begin{abstract}
We study the emission properties of thermonuclear explosions in a helium white dwarf (WD) tidal disruption event (TDE).
We consider a TDE where a 0.2~$M_{\sun}$ helium WD is disrupted by a $10^{2.5}\,M_{\sun}$ intermediate-mass black hole (IMBH).
The helium WD is not only tidally disrupted but is also detonated by the tidal compression and by succeeding shocks.
We focus on the emission powered by radioactive nuclei in the unbound TDE ejecta.
We perform hydrodynamic simulations coupled with nuclear reactions, post-process detailed nucleosynthesis calculations, and then radiative transfer simulations.
We thus derive multi-band light curves and spectra.
The helium WD TDE shows rapid ($\Delta t_{1\mathrm{mag}}\simeq5\text{--}10$~days) and relatively faint ($L_{\mathrm{peak}}\simeq10^{42}\,\mathrm{erg}\,\mathrm{s}^{-1}$) light curves, because the ejecta mass and $^{56}$Ni mass are low ($0.12\,\mathrm{M}_{\sun}$ and $0.03\,\mathrm{M}_{\sun}$, respectively).
The spectra show strong calcium and Fe-peak features and very weak silicon features, reflecting the peculiar elemental abundance.
    The key feature is the Doppler shift of the spectral lines up to $\simeq\pm12,000\,\mathrm{km}\,\mathrm{s}^{-1}$, depending on the viewing angle, due to the bulk motion of the ejecta.
Our model matches well with two rapid and faint transients reported in \citet{2018MNRAS.481..894P}.
The particular model presented here does not match with observed SNe Iax, calcium-rich transients, or .Ia explosion candidates, either in the spectra or light curves.
However, we expect a large variety of the observational signatures once a wide range of the WD/black hole masses and orbital parameters are considered.
This study contributes to the search for WD TDEs with current and upcoming surveys, and to the identification of IMBHs as disrupters in the TDEs.
\end{abstract}


\section{Introduction}
\label{sec:intro}
In a tidal disruption event (TDE), a star is tidally disrupted by a black hole (BH) when the distance between them is smaller than a tidal radius, $R_t \equiv R_\star (\Mbh / M_\star)^{1/3}$ (e.g. \citealt{1988Natur.333..523R}).
After the disruption, about half of the debris is bound to the BH, leading to emission across a wide wavelength range (for a review, see \citealt{2019GReGr..51...30S}).

% descrive WD TDE
% nuclaer, IMBH
If the disrupted star is a white dwarf (WD), only an intermediate-mass BH (IMBH) with $\lesssim 10^5\,\Msun$ can disrupt it.
By contrast, a supermassive BH (SMBH) with $\gtrsim 10^5\,\Msun$ cannot tidally disrupt a WD because the SMBH swallows the WD before the tidal disruption \citep{1989A&A...209..103L}.
Therefore, the WD TDEs are useful for studying the existence and properties of IMBHs.

The WD TDEs have another unique feature: 
the WD is not only tidally disrupted but also can be detonated by the tidal compression and by succeeding shocks.
The thermonuclear explosions triggered by the detonation occur if the encounter is sufficiently deep inside the tidal radius (e.g. \citealt{1982Natur.296..211C,2009ApJ...695..404R}).
% \citep{1982Natur.296..211C,1983ApJ...273..749B,1989A&A...209...85L,2008ApJ...679.1385R,2009ApJ...695..404R,2012ApJ...749..117H,2015MNRAS.450.4198S,2017ApJ...839...81T}.
% \citep{2017ApJ...839...81T,2018MNRAS.475L..67T,2018MNRAS.477.3449K,2018ApJ...858...26T}
It releases nuclear energy, affects dynamics of the WD debris, and produces radioactive nuclei that could later power the emission from the unbound ejecta.

% LSST, ZTF
% Holcomb+ Sell+ Ca-rich
\citet{2016ApJ...819....3M} studied the properties of the emission from thermonuclear explosions in a carbon-oxygen (CO) WD TDE.
They showed that the emission is reminiscent of supernovae (SNe) I.
% Still, we do not know a variety of observational signatures of thermonuclear emission from WD TDEs because \citet{2016ApJ...819....3M} only consider one parameter set case of CO WD TDE.
However, \citet{2016ApJ...819....3M} only considered one particular case of a CO WD TDE.
Therefore, it is not clear if other WD TDEs share the same properties shown in their model once other parameter sets (e.g. different WD/BH masses) are considered.

WD TDEs are interesting targets for current and upcoming transient surveys.
\citet{2016ApJ...819....3M} also estimated the detection rate of the thermonuclear emission from WD TDEs.
They showed that the Vera C. Rubin Observatory (previously known as the Large Synoptic Survey Telescope) may be able to detect tens of the events per year if a number density of IMBHs is $\simeq0.02\,\mathrm{Mpc}^{-3}$, which is still unknown.
Even current surveys such as the Zwicky Transient Facility (ZTF) may be able to detect these events if the number density of IMBHs is larger.
It is useful to derive synthetic light curves and spectra of WD TDEs in order to search for them with such surveys.

Here, we study properties of the emission from a WD TDE where an IMBH disrupts a helium WD.
A motivation for this is that we can intuitively expect different observational signatures of helium WD TDEs from those of CO WD TDEs; the ejecta mass and nucleosynthetic yields, such as $\nickel$ mass, significantly differ between the two types \citep{2018MNRAS.477.3449K}.
Another motivation is that helium WD TDEs have been suggested to be an origin of calcium-rich transients \citep{2015MNRAS.450.4198S,2018MNRAS.475L.111S}.
\citet{2015MNRAS.450.4198S} proposed that the properties of the calcium-rich transients may be explained by the helium WD TDE, based on hydrodynamic models by \citet{2009ApJ...695..404R}.
However, the models by \citet{2009ApJ...695..404R} lack detailed elemental abundance and synthetic observations.
Thus, it is still unclear whether helium WD TDEs can be the origin of calcium-rich transients (see also \citealt{2019ApJ...887..180S}).
% In this study, we overcome the difficulties by performing detailed nucleosynthesis calculations and radiative transfer simulations.
In this study, we perform detailed nucleosynthesis calculations and radiative transfer simulations, and predict synthetic observables for the helium WD TDE.

%%%%%%%%%%%%%%%%%%%%%%
%%%%% structure  %%%%%
%%%%% not needed %%%%%
%%%%%%%%%%%%%%%%%%%%%%

The structure of this Letter is as follows.
We describe our numerical methods and hydrodynamical models in \secref{sec:method}.
We present synthetic multi-band light curves and spectra in \secref{sec:result}.
% We discuss the results in \secref{sec:discussion}.
Discussions are given in \secref{sec:discussion}, where our model is compared to observational properties of some transients whose origins are not yet clarified.

\section{Methods and models}
\label{sec:method}

We consider a helium WD TDE with a parameter set of $\Mbh = 10^{2.5}\,\Msun$, $\Mwd=0.2\,\Msun$, and a penetration parameter $\beta \equiv R_t/R_p =5$, where $R_p$ is a pericenter radius of an orbit, with a pure $^4$He composition of the WD.
Additionally, we consider a CO WD TDE with the parameter set used in \citet{2016ApJ...819....3M}, $\Mbh=500\,\Msun$, $\Mwd=0.6\,\Msun$, and $\beta=5$ with a composition of 50~\% mass fractions of $^{12}$C and $^{16}$O, in order to cross-check our methods with theirs.
% We use three steps of numerical simulations.

\subsection{Hydrodynamic Simulations}
First, we perform smoothed particle hydrodynamic (SPH) simulations coupled with nuclear reactions.
We summarize the methods here; detailed description can be found in \citet{2017ApJ...839...81T} and \citet{2018MNRAS.477.3449K}.
% (see also \citealt{2015ApJ...807..105S,2016ApJ...821...67S}).
The initial separation between the BH and WD is $5 R_t$.
We use \texttt{Helmholtz} equation of state with Coulomb correction \citep{2000ApJS..126..501T}.
The SPH simulations are coupled with $\alpha$-chain nuclear reaction networks among 13 species \citep{2000ApJS..129..377T}.
% We take into account the self-gravity of the WD \citep{1997JCoPh.136..298M}.
We use the gravitational potential of \citet{2013MNRAS.433.1930T} to include an approximate general relativistic correction for the Schwarzschild BH.
For the CO WD TDE, we apply the Newtonian gravity in order to set the same condition as that of \citet{2016ApJ...819....3M}.
% We eliminate SPH particles when a distance between a SPH particle and the center of the BH is shorter than than the sum of the kernel support radius of the SPH particle and the Schwarzschild radius.
We terminate the simulation at 2000~s (1000~s) for the helium (CO) WD TDE, when the homologous expansion of the ejecta is well realized.

We employ 786,432 (393,216) SPH particles to represent the helium (CO) WD on a parabolic orbit around the BH.
These resolutions are not enough to resolve shock structure during the tidal compression, and thus the nucleosynthetic results are resolution dependent \citep{2017ApJ...839...81T}.
% \citet{2017ApJ...839...81T} show that nucleosynthetic results of three-dimensional hydrodynamic simulations of WD TDEs do not converge even when they use about 25 million SPH particles because they fail to resolve shock structure during the tidal compression.
% We use basically same methods as used in \citet{2017ApJ...839...81T} and we use less SPH particles than \citet{2017ApJ...839...81T}.
% Therefore, our results are affected by numerical resolutions.
However, we expect that our results do not change significantly.
One reason is that detonations in helium WD TDEs are confirmed in \citet{2018ApJ...858...26T} and in \citet{2018ApJ...865....3A}, who use independent numerical methods.
Another reason is that we also see rough matches between nucleosynthetic yields of our hydrodynamic simulations (see \citealt{2018MNRAS.477.3449K}) and those of \citet{2018ApJ...865....3A}.
% Furthermore, we expect large variety of thermonuclear emission from WD TDEs for other parameter cases.
% Even if our nucleosynthetic results are not exactly correct, similar nucleosynthesis and succeeding thermonuclear emission would be given for other parameter cases.

\begin{figure}
    \plotone{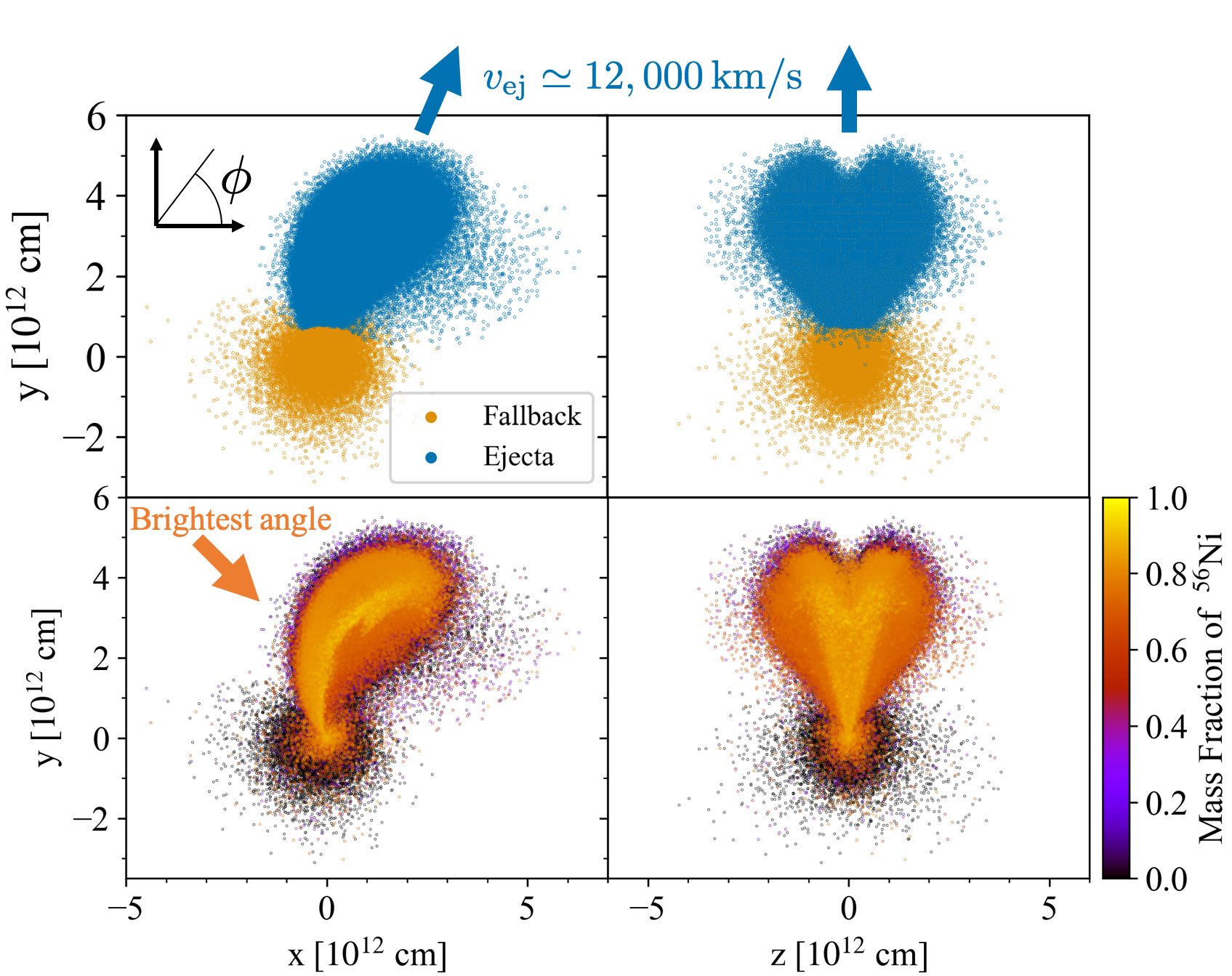}
    \caption{Distribution of helium WD TDE debris at the end of the hydrodynamic simulation.
    The upper panels show the distribution of fallback and ejecta debris, and the lower panels show the $\nickel$ distribution.
    % We also show the azimuthal angle $\phi$ of the spherical coordinates used in the radiateive transfer simualation.
    }
    \label{fig:distribution}
\end{figure}

\figref{fig:distribution} shows the distribution of the fallback/ejecta debris and $\nickel$ at the end of the helium WD TDE simulation.
The shape of the ejecta is very aspherical due to the tidal disruption.
The properties of the unbound ejecta are as follows.
The mass is $0.12\,\Msun$.
Its center of mass (COM) escapes from the BH with the velocity of $12,000\,\kmpersec$.
The kinetic energy with respect to the COM is $6.5\times10^{49}$~erg.
Because we focus on the emission from the unbound ejecta, the fallback debris, which is bound to the BH after the first approach, is hereafter ignored.

\subsection{Detailed Nucleosynthesis Calculations}
\label{sec:nucleosynthesis}

Second, we perform detailed nucleosynthesis calculations to derive accurate and detailed elemental abundances of the ejecta with the \texttt{torch} code\footnote{\url{http://cococubed.asu.edu/}} \citep{2000ApJS..129..377T}.
We record histories of density and temperature for all the SPH particles in the hydrodynamic simulations during phases when explosive nuclear reactions take place.
Then we perform the nucleosynthesis calculations for all the particles in a post-process manner with the density-temperature histories, considering networks among 640 isotopes.
We take the same initial nuclear composition as that those used in the SPH simulations.

\begin{figure}
    \plotone{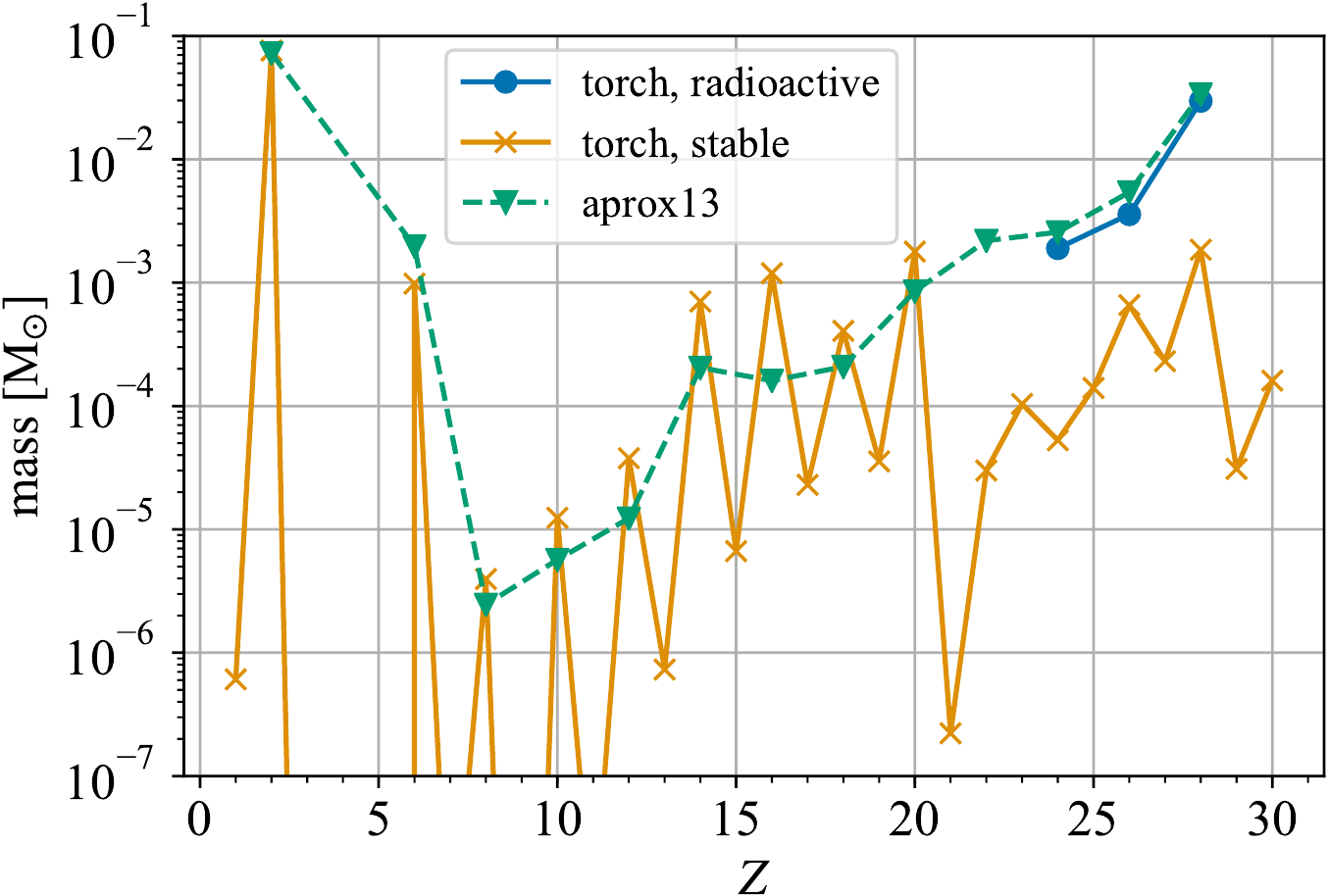}
    \caption{Nucleosynthesis in our helium WD TDE case.
    The blue and orange lines respectively show the masses of radioactive and stable elements in the unbound ejecta derived from the post-process detailed nucleosynthesis calculations.
    The green line shows those derived from the simple nuclear reaction networks adopted in the hydrodynamic simulation.}
    \label{fig:abundance}
\end{figure}

\figref{fig:abundance} shows the elemental abundance of the ejecta in the helium WD TDE, which is derived from this method.
For comparison, we also show abundance derived from the simplified network adopted in the hydrodynamic simulations.
$\nickel$ is dominant in the radioactive nuclei with its mass of $0.030\,\Msun$.
$\calcium$ is dominant in the intermediate-mass elements (IMEs) with the mass of $0.0014\,\Msun$, while $\silicon$ is sub-dominant with its mass of $7.0 \times 10^{-4}\,\Msun$.
Interestingly, this abundance pattern is quite different from SNe Ia and from a CO WD TDE, while it is qualitatively consistent with previous studies on helium WD detonations \citep{2013ApJ...771...14H}.

\subsection{Radiative Transfer Simulations}

We perform radiative transfer simulations using \texttt{HEIMDALL} \citep{2006ApJ...644..385M,2014ApJ...794...37M}.
%% describe HEIMDALL here
\texttt{HEIMDALL} can handle multi-dimensional, multi-frequency, and time-dependent Monte Carlo radiative transfer simulations.
We assume local thermal equilibrium and homologous expansion of the ejecta.
We consider stable elements with atomic numbers $Z$ from 1 to 30.
Radioactive decays of $^{48}$Cr/V/Ti, $^{52}$Fe/Mn/Cr, and $\nickel$/Co/Fe are included as the power sources.
The computational domain is set such that its origin is at the COM of the ejecta.
The three-dimensional spherical grid of $(v_r, \cos \theta, \phi)$ is equally sampled with bins of $(100, 50, 50)$, where the outer edge of the radial grid is set at the velocity of $2 \times 10^4\,\kmpersec$, corresponding to the radius of $4\times10^{12}$~cm at the end of the SPH simulation (see \figref{fig:distribution}).
We map the results of the SPH simulations and detailed nucleosynthesis calculations to the grid following the prescription by \citet{2018arXiv180303652R}, which can well conserve integral fluid properties and can maintain resolution of the SPH data.
% We perform the radiative transfer simulations from 3~\d to 70\~days after the WD disruption.
Note that results of the helium WD TDE in a later phase, after $\simeq30$~days, are uncertain because the ejecta become optically thin and thus the local thermal equilibrium assumption becomes invalid.

\section{Results: light curves and spectra}
\label{sec:result}

% \subsection{Light curves}
% \label{sec:lightcurve}

\begin{figure}
    \plotone{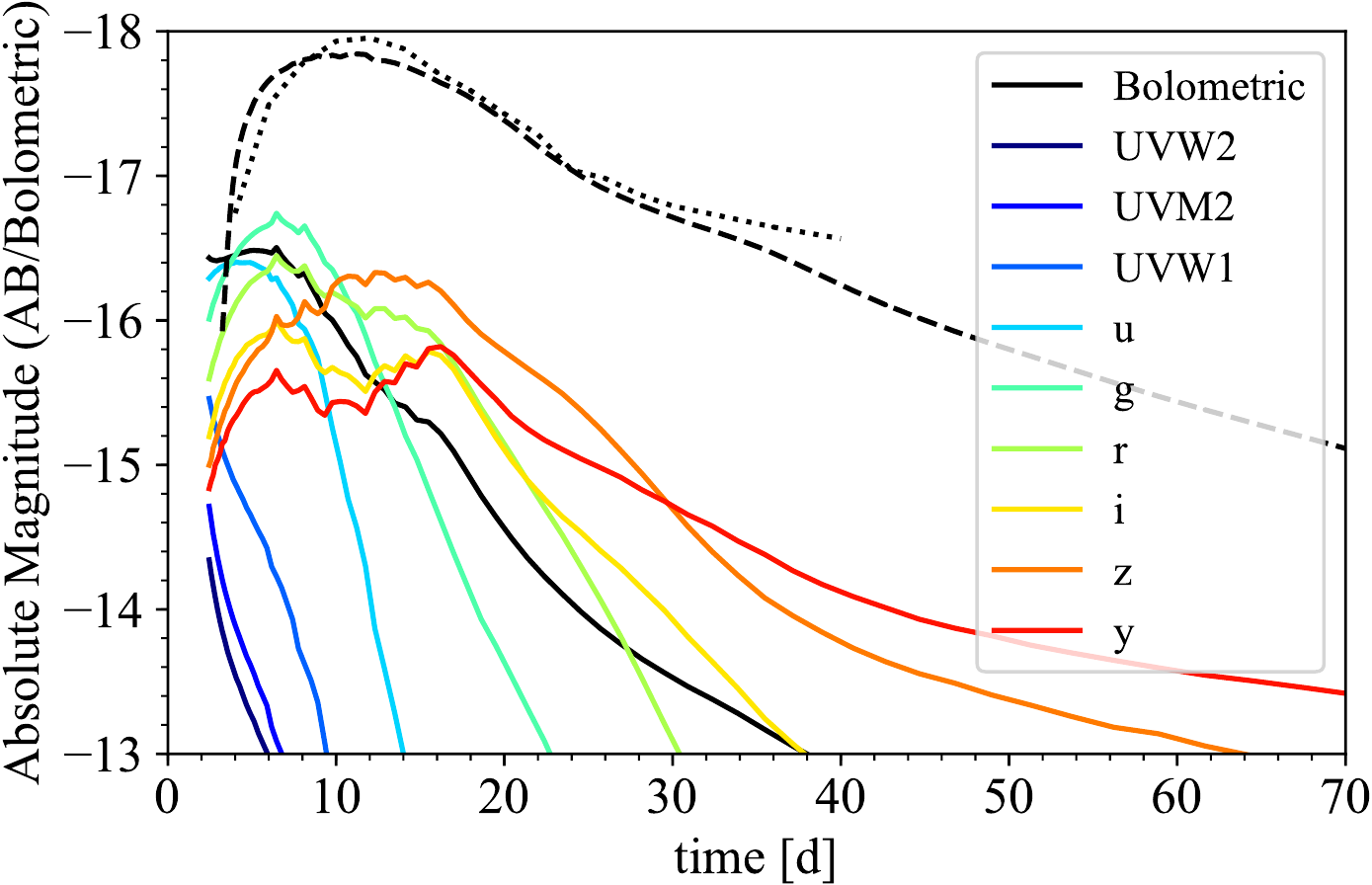}
    \plotone{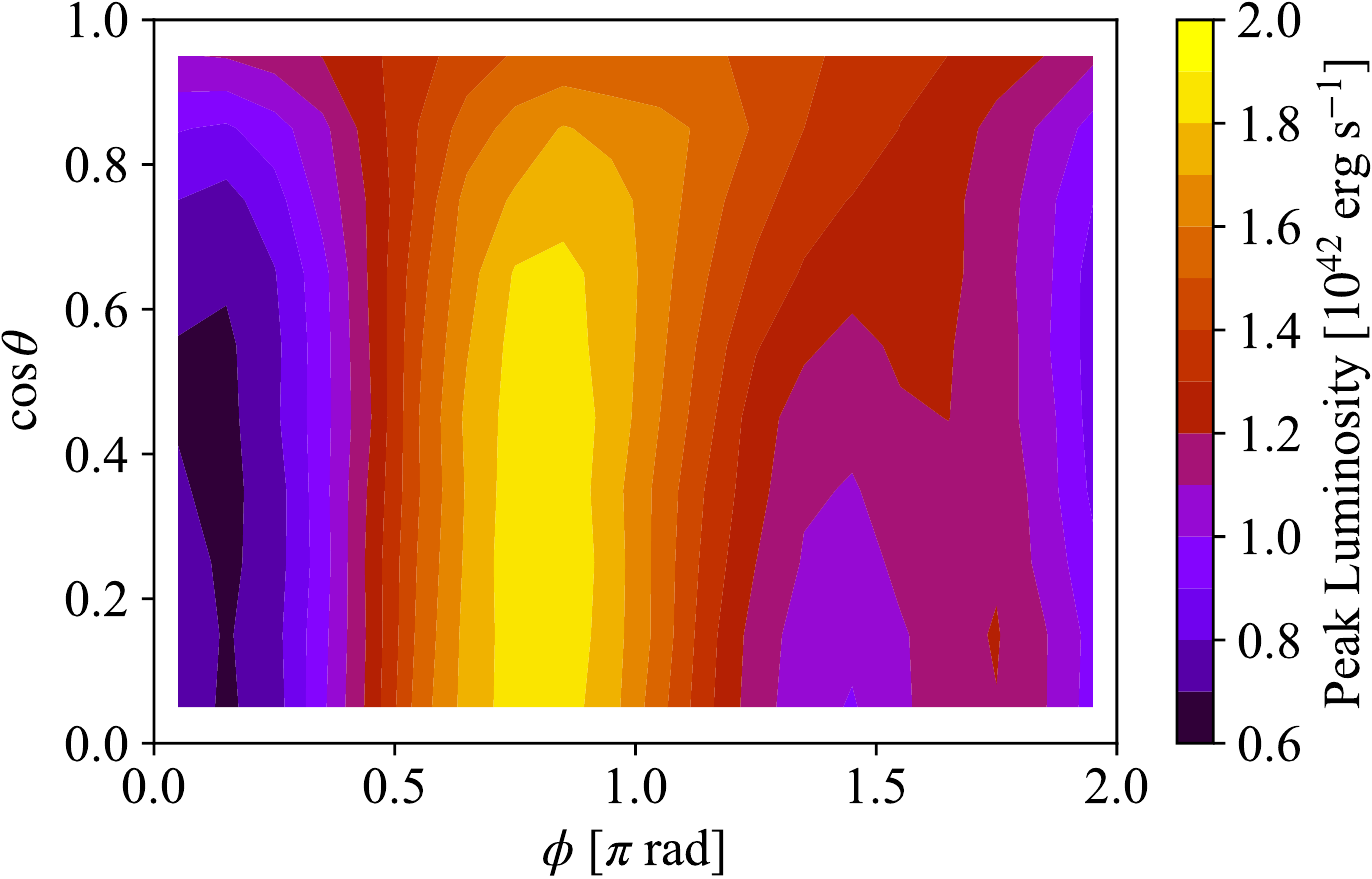}
    \caption{Top panel: light curves of WD TDEs. The solid curves show the bolometric and multi-band light curves of the helium WD TDE.
    The black dashed and dotted curves, respectively, show the bolometric light curves of the CO WD TDE derived in this study and in \citet{2016ApJ...819....3M}.
    Bottom panel: isotropic equivalent peak luminosity as a function of viewing angle.}
    \label{fig:lightcurve}
\end{figure}

The top panel of \figref{fig:lightcurve} shows our synthetic multi-band light curves of the helium WD TDE.
We consider filters of the Swift and Vera C. Rubin Observatory here.
The origin of the time is when the first WD-BH pericenter passage occurs.
The top panel of \figref{fig:lightcurve} also compares bolometric light curves of the CO WD TDE derived in this study and in \citet{2016ApJ...819....3M},\footnote{The data are available on \url{https://github.com/morganemacleod/WhiteDwarf_Thermonuclear_Transients}} in order to cross-check the methods employed in this work.
We see a good agreement for the CO WD TDE models.
% It is clearly shown that both the CO WD TDE models are in good agreement, indicating validity of our calculations.
Hereafter, we will focus on the helium WD TDE.

The helium WD TDE appears fainter, with more rapid evolution than the CO WD TDE.
Depending on the viewing angle, the isotropic equivalent peak luminosity is $L_{\mathrm{peak}} \simeq 0.7\text{--}2.0 \times 10^{42}\,\ergpersec$ (see the bottom panel of \figref{fig:lightcurve}).
The time in which the bolometric luminosity declines by one magnitude from the peak is $\Delta t_{1\mathrm{mag}} \simeq 5\text{--}10$~days.
These features are attributed to a less-massive ejecta mass ($0.12\,\Msun$) and $\nickel$ mass ($0.030\,\Msun$) than the CO WD TDE.
% todo put diffusion timescale equation
% todo put peak luminosity equation
% These luminosities and timescales are similar to those of calcium-rich transients \citep{2010Natur.465..322P,2012ApJ...755..161K,2014MNRAS.437.1519V,2017ApJ...836...60L,2017ApJ...846...50M,2018ApJ...866...72D}, .Ia explosions \citep{2010ApJ...715..767S,2010ApJ...723L..98K}, and some rapid transients recently discovered by \citet{2014ApJ...794...23D,2018MNRAS.481..894P}.
The luminosities and timescales are similar to those of calcium-rich transients, .Ia explosions, and rapid transients (see \secref{sec:obs}).
The color of the light curves also show rapid evolution: $g-r\simeq-0.4$~mag and $r-i\simeq-0.5$~mag around the peak, while $g-r\simeq+0.9$~mag and $r-i\simeq-0.2$~mag at 10~days after the peak.
The near-infrared (NIR) $izy$ bands show the second hump at $t\simeq12\text{--}18$~days, which is commonly seen in SNe Ia, and originates from the recombination of Fe/Co III to Fe/Co II \citep{2006ApJ...649..939K}.

\begin{figure*}
    \gridline{
        \fig{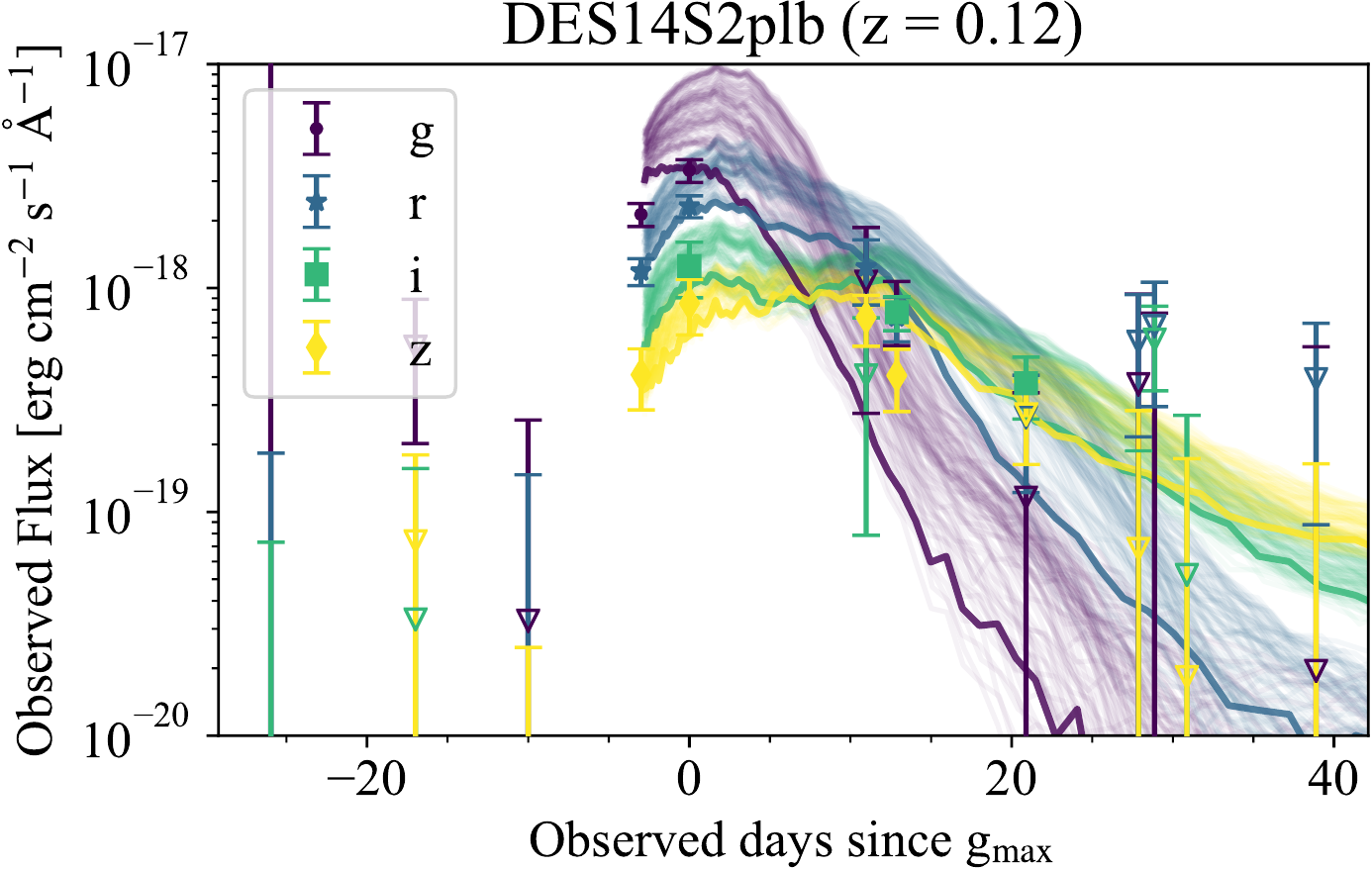}{0.33\textwidth}{(a) DES14S2plb (rapid and faint)}
        \fig{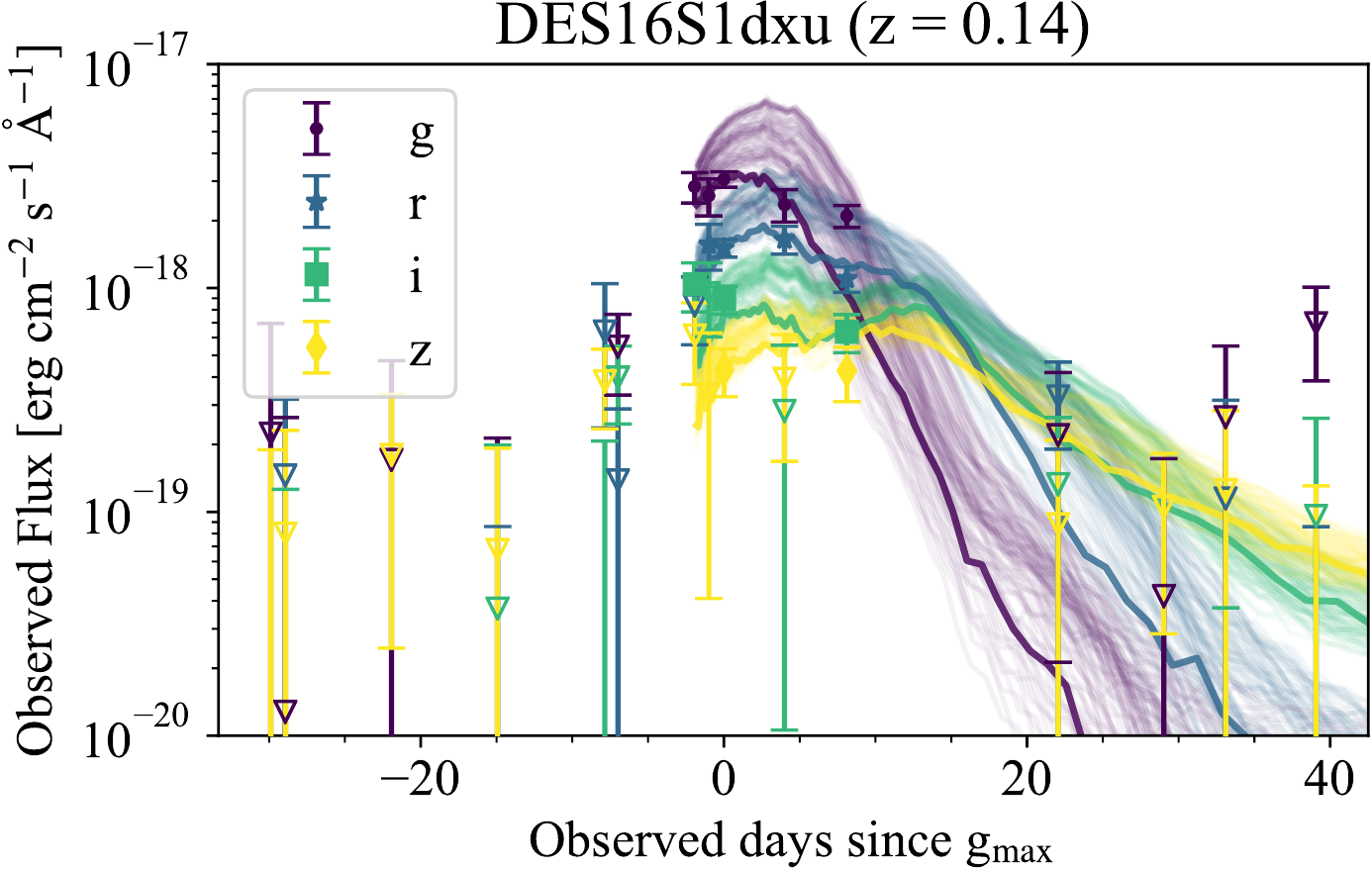}{0.33\textwidth}{(b) DES16S1dxu (rapid and faint)}
        \fig{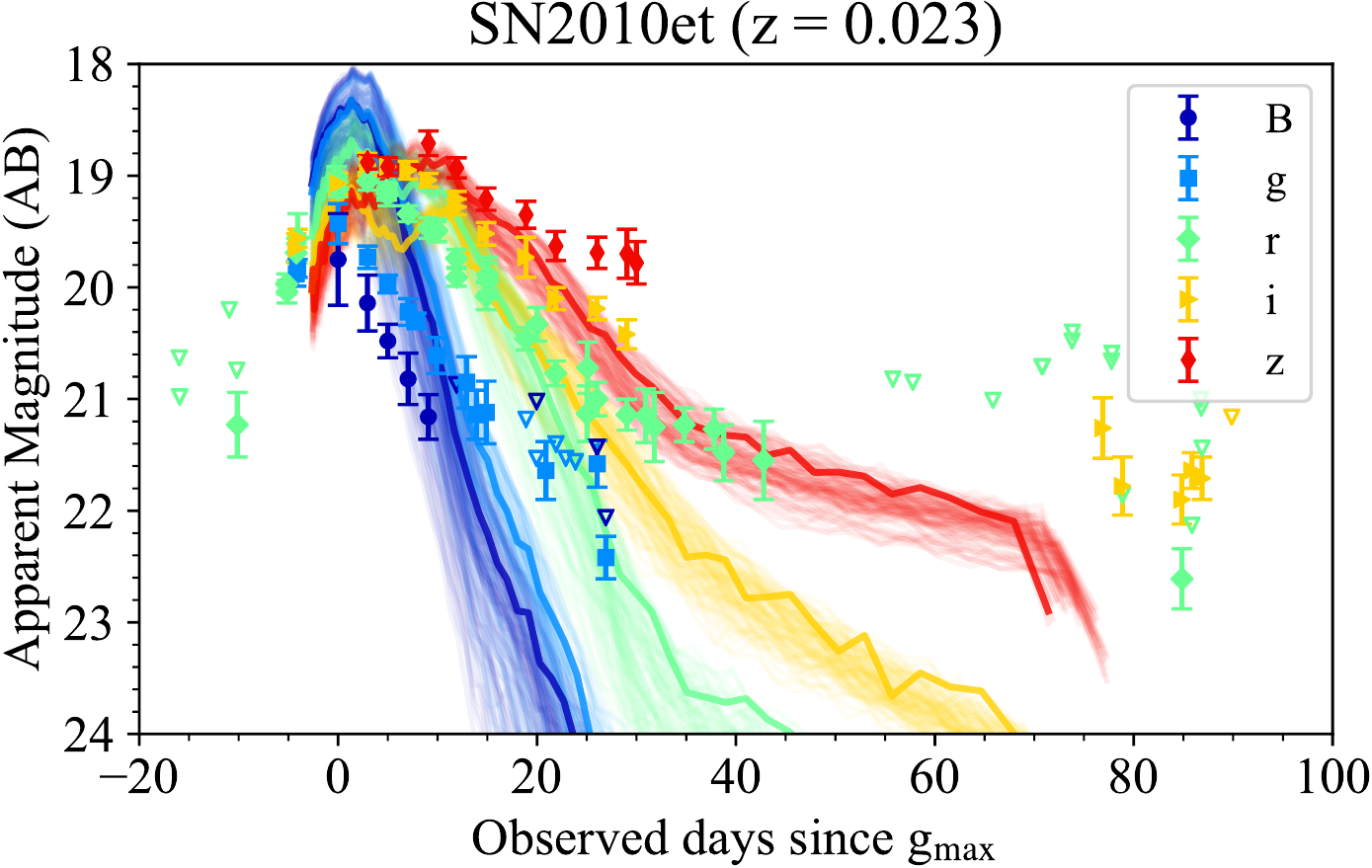}{0.33\textwidth}{(c) SN2010et (calcium-rich)}
    }
    \gridline{
        \fig{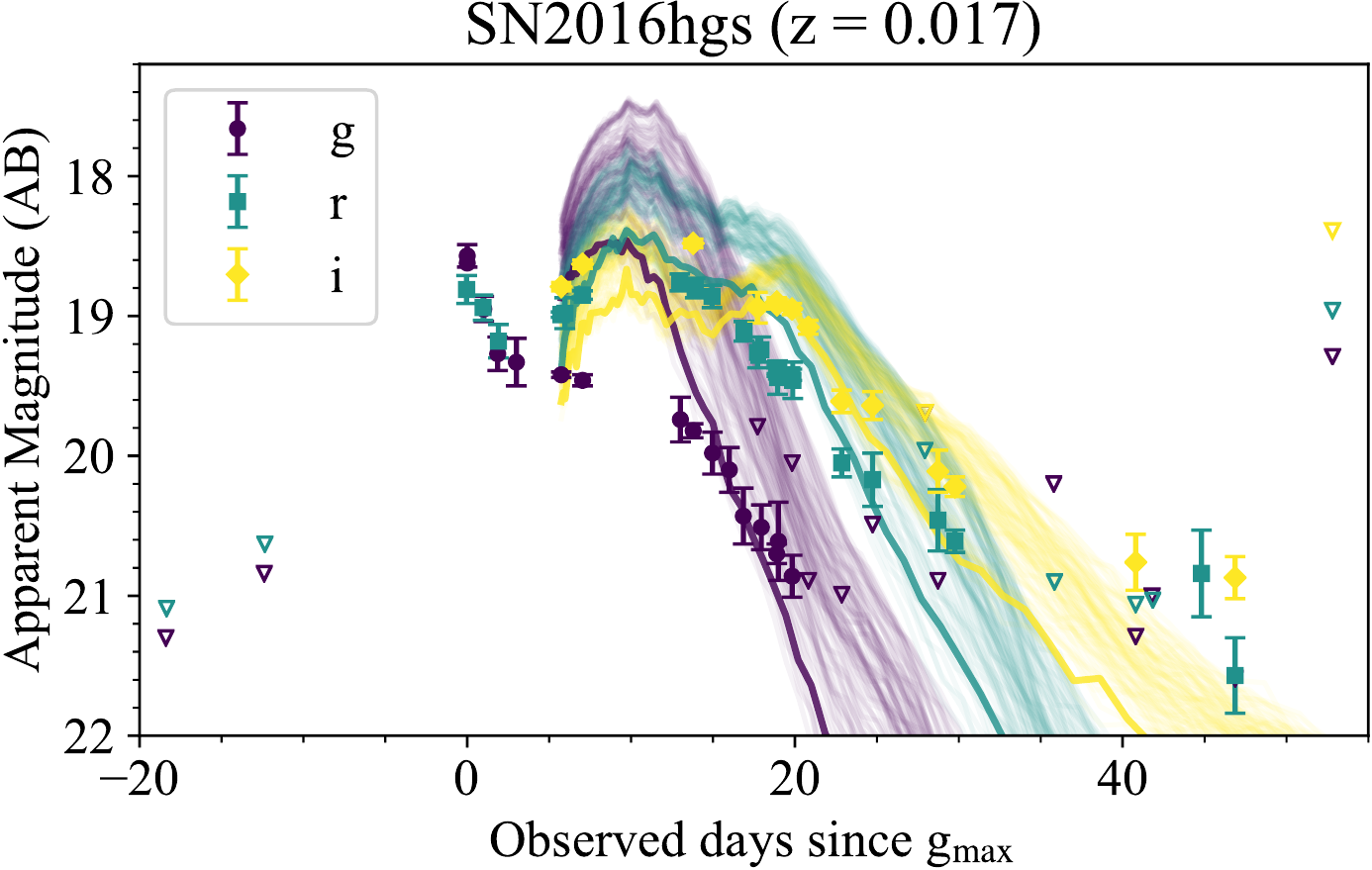}{0.33\textwidth}{(d) SN2016hgs (calcium-rich)}
        \fig{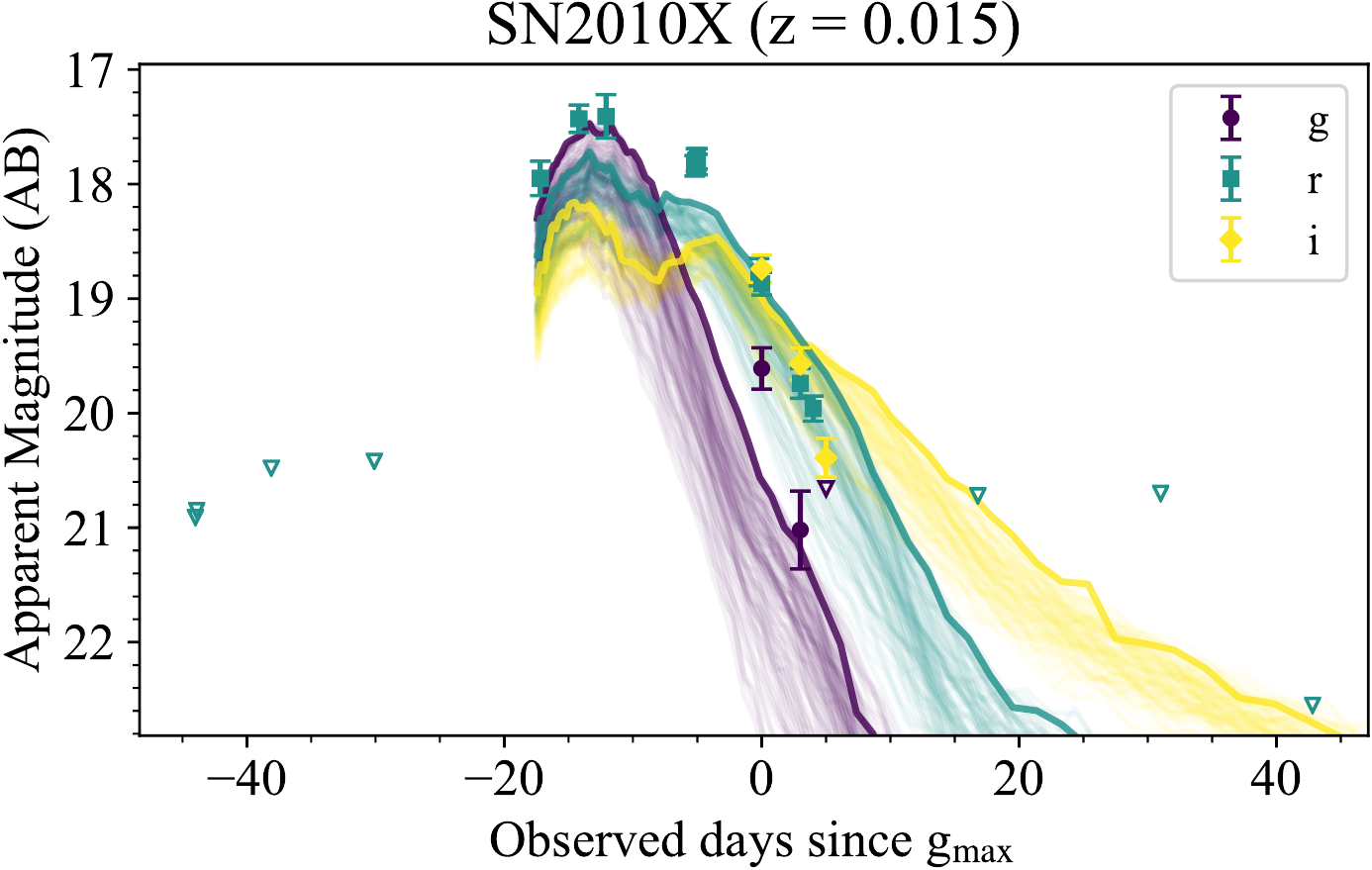}{0.33\textwidth}{(e) SN2010X (.Ia explosion candidate)}
        \fig{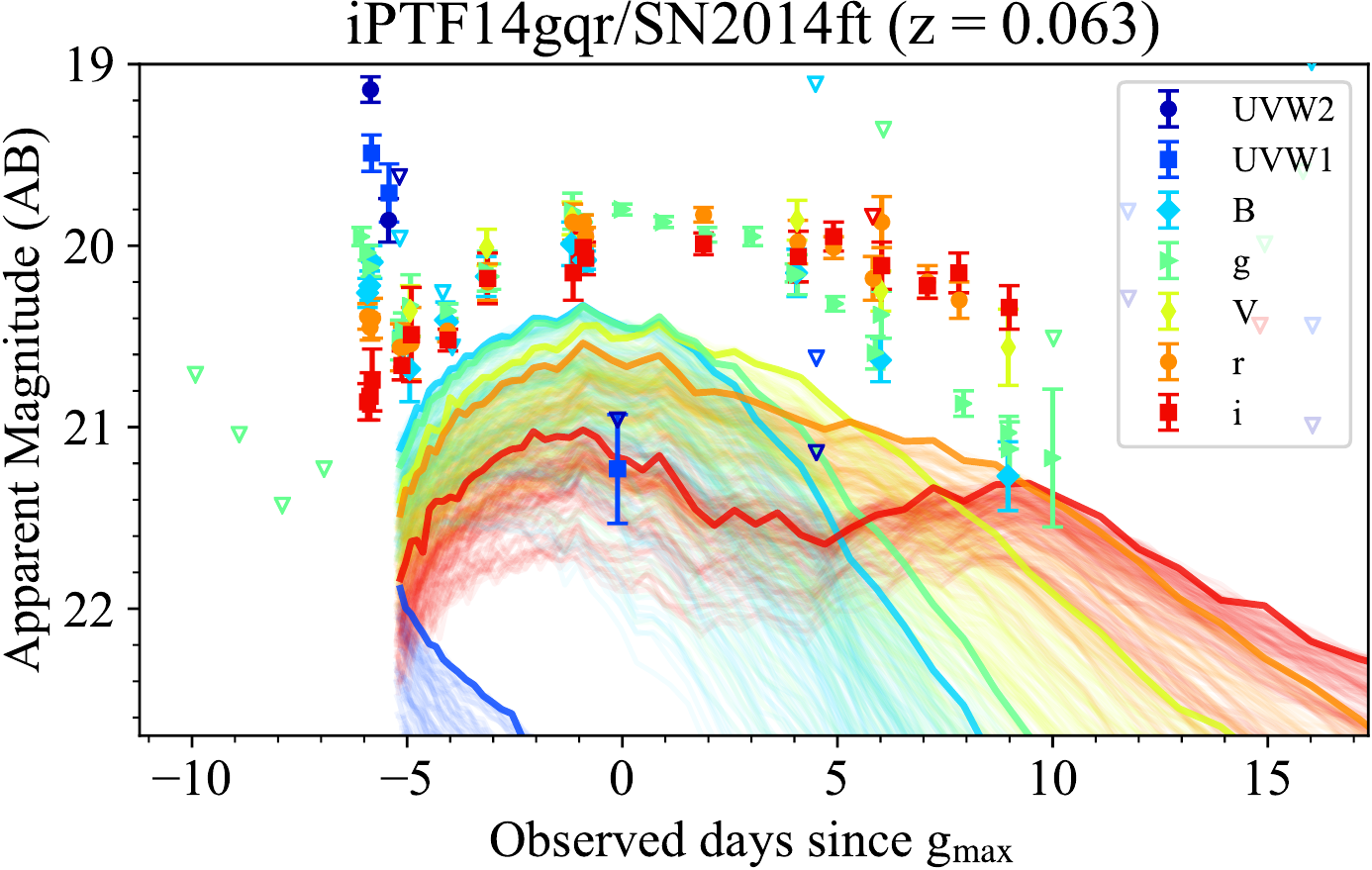}{0.33\textwidth}{(f) iPTF14gqr/SN2014ft (rapid Type Ic)}
    }
    \gridline{
        \fig{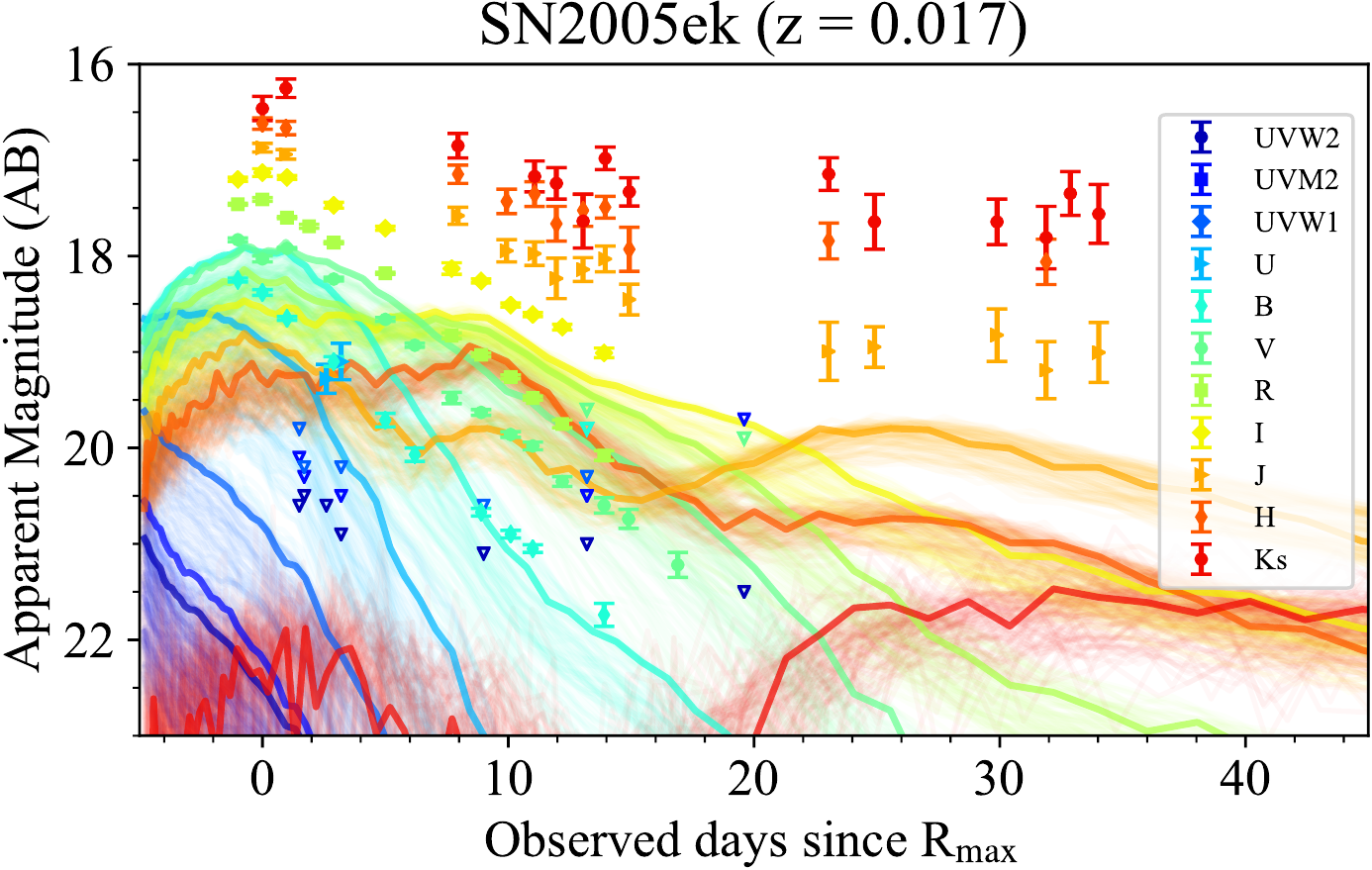}{0.33\textwidth}{(g) SN2005ek (rapid Type Ic)}
    }
    % \plottwo{}
    \caption{Comparisons between our model and some rapid transients reported so far.
    We compare them with rapid and faint transients reported in \citet{2018MNRAS.481..894P}, calcium-rich transients SN2010et \citep{2012ApJ...755..161K} and SN2016hgs \citep{2018ApJ...866...72D}, a .Ia explosion candidate SN2010X \citep{2010ApJ...723L..98K}, and rapid Type Ic transients iPTF14gqr/SN2014ft \citep{2018Sci...362..201D} and SN2005ek \citep{2013ApJ...774...58D}.
    The points show the observed flux/magnitude and its 1$\sigma$ error.
    The open triangles in the panels (a, b) are the cases where the detection significance is less than 3$\sigma$, while those in the other panels show upper limits.
    The thin solid curves show our model light curves with different viewing angles $(\cos\theta, \phi)$ equally sampled with $(10, 10)$ bins.
    The thick solid curves show those with the angle where the model can best fit the observations.
    }
    \label{fig:obs}
    % \plottwo{figs/DES14S2plb.pdf}{figs/DES16S1dxu.pdf}
    % % \plottwo{}
    % \caption{Comparisons between our models and rapid transients reported in \citet{2018MNRAS.481..894P}.
    % The points show the observed flux and its 1$\sigma$ error.
    % The open triangles are the cases where the detection significance is less than 3$\sigma$.
    % The solid curves show our model light curves with different viewing angles.
    % The thick curves show those with the angle where the model can fit the observed light curves best.
    % }
    % \label{fig:obs}
\end{figure*}

The bottom panel of \figref{fig:lightcurve} shows a strong dependence of the peak luminosity on the viewing angle.
\figref{fig:obs} also shows how the multi-band light curves vary by viewing angle.
The figure also shows comparisons to a few observed transients, which will be discussed in \secref{sec:discussion}.
The viewing angle dependence reflects the aspherical distribution of the ejecta (see \figref{fig:distribution}), which is a unique feature of the tidal disruption by the BH.
% The strong dependence on the viewing angle shown in the bottom panel of \figref{fig:lightcurve} reflects the aspherical distribution of the ejecta (see \figref{fig:distribution}), which is a unique feature of the tidal disruption by the BH.
% \figref{fig:obs} shows how the multi-band light curves vary by the viewing angles.
% The figure also shows comparisons to a few observed transients, which will be discussed in \secref{sec:discussion}.
The brightest emission is realized for a viewing angle $\phi\simeq0.8\pi$, where a projected surface area of the ejecta is maximized and also photons escape to the ejecta surface more quickly than at other angles.
The short photon diffusion timescale leads to a fast decline ($\Delta t_{1\mathrm{mag}}\simeq5$~days) and bluer color ($g-r\simeq-0.5$ around the peak) for this brightest viewing angle.
By contrast, the faintest event is the case for $\phi\simeq0.2\pi$, which shows a slow decline ($\Delta t_{1\mathrm{mag}}\simeq10$~days) and redder color ($g-r\simeq-0.1$ around the peak).

These viewing angle dependences are qualitatively consistent with the CO WD TDE model of \citet{2016ApJ...819....3M}, although the peak luminosity of our model varies by a factor of $\simeq3$, while that of \citet{2016ApJ...819....3M} varies by a factor of $\simeq10$.
The difference arises because the ejecta shape of our model is less distorted from the spherical than that of \citet{2016ApJ...819....3M}.
This difference would generate a different behavior of the NIR second hump: contrary to our helium WD TDE model, the CO WD TDE model of \citet{2016ApJ...819....3M} does not show the NIR second hump.
The second hump is produced when the NIR emissivity of Fe/Co sharply increases at a temperature of $\simeq7000$~K because of recombination \citep{2006ApJ...649..939K}.
The ejecta of the CO WD TDE model is more aspherical and thus the photosphere temperature varies significantly depending on the viewing angle (see Figure~7 in \citealt{2016ApJ...819....3M}).
At a particular moment, only a small part of photosphere would be at the recombination temperature and thus would contribute to the NIR emission.
Photon scattering would mix photons from different angles with different temperatures, and then the NIR emission by the recombination would be smoothed out.
As a result, the second hump in the CO WD TDE would be much weaker than the helium WD TDE.

% We also expect that thermonuclear emission from WD TDEs may show polarization because the shape of the ejecta is very aspherical \citep{1982ApJ...263..902S,1991A&A...246..481H,1996ApJ...459..307H}.
We expect that thermonuclear emission from WD TDEs may show polarization because of the aspherical shape of the ejecta \citep{1982ApJ...263..902S,1991A&A...246..481H}.
We plan to study this issue in our future work.

% \subsection{Spectra}
% \label{sec:spectra}

\begin{figure*}
    \plottwo{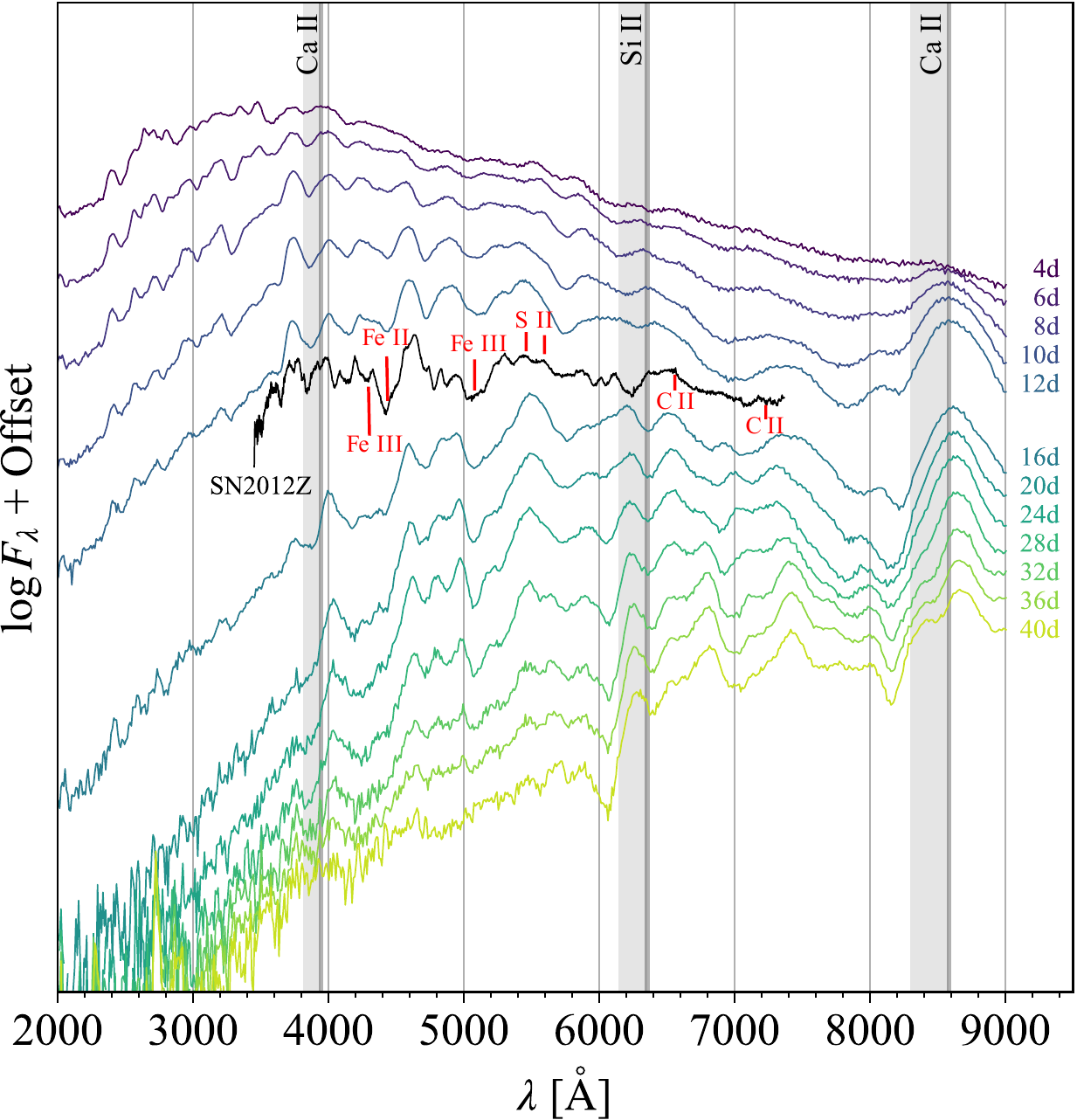}{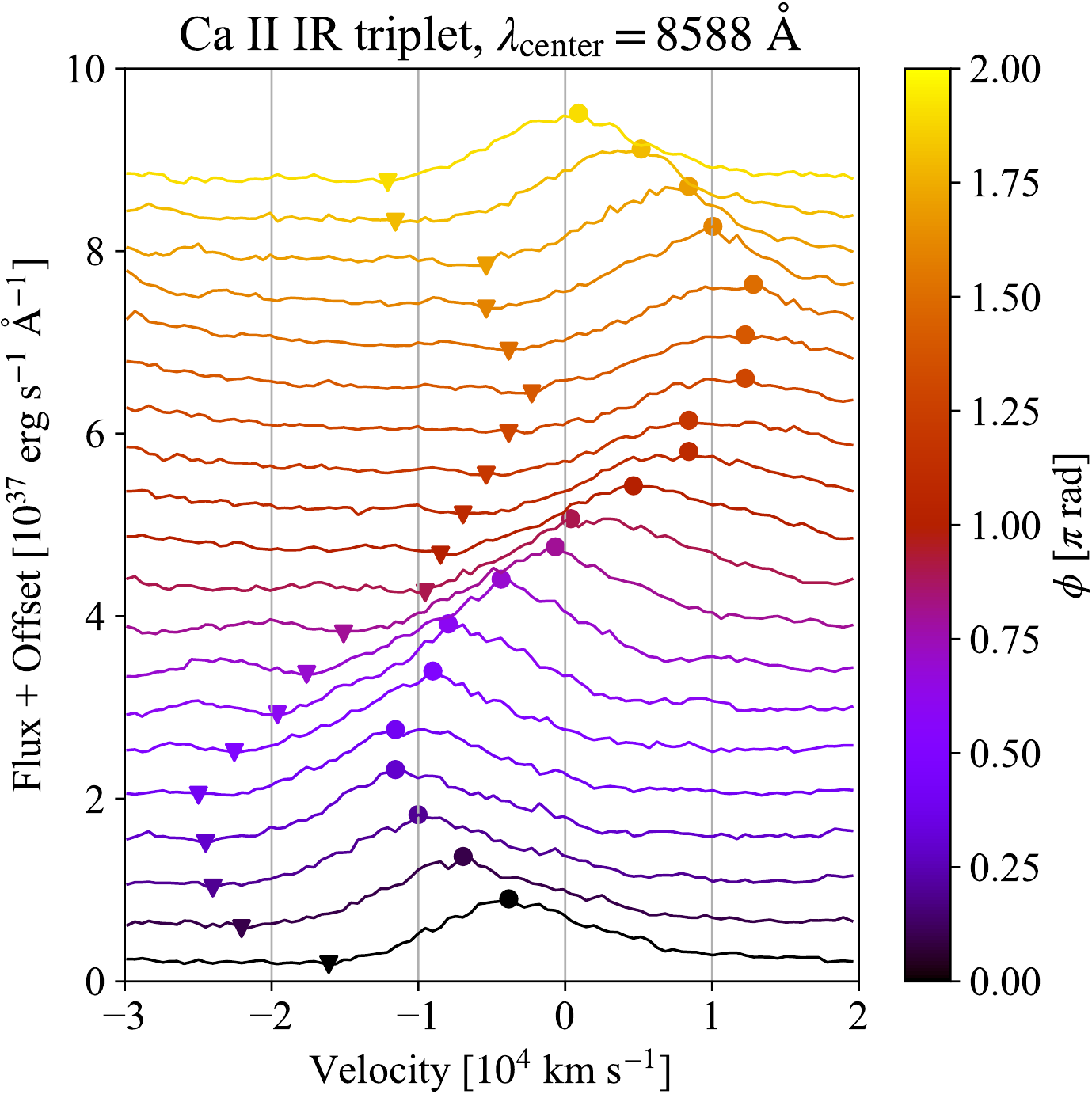}
    \caption{Time evolution and viewing-angle dependence of spectra of the helium WD TDE.
    The left panel shows the spectral evolution, which is calculated by taking the mean over all the angles in the ejecta-comoving frame.
    The vertical thick gray lines show rest-frame wavelengths of some lines seen in SNe Ia, and the vertical gray shades show ranges of the wavelengths shifted with a velocity range from $-10,000\,\kmpersec$ to $0\,\kmpersec$.
    Some other spectral lines are shown with the red labels.
    For comparison, we also show a $+1.1$~days spectrum of SN2012Z since its $V$ maximum with the black line \citep{2013ApJ...767...57F}.
    The right panel shows the viewing angle dependence of the shape of Ca II IR triplet lines at $t \simeq 16$~days.
    The circles show peaks of the emission lines, and the triangles show the minima of the absorption lines.
    We take the mean over the angle of $-0.3 < \cos \,\theta < 0.3$ with 20 samples of $\phi$.
    }
    \label{fig:spectra}
\end{figure*}

The left panel of \figref{fig:spectra} shows the time evolution the synthetic spectra of the helium WD TDE and a spectrum of SN Iax 2012Z, which shows similar features to the helium WD TDE (we will compare them in \secref{sec:obs} ).
They share some common features with SNe Ia: the lack of hydrogen lines, the appearance of strong Fe II lines from $4000$ to $5000\,\mathrm{\AA}$, and P Cygni profiles.
Because of the abundance of IMEs in the helium WD TDE, strong Ca II H/K and infrared (IR) triplet emission lines appear, and silicon emission/absorption lines around $6150\,\mathrm{\AA}$ are, interestingly, absent or very weak, depending on the viewing angle.
The appearance/lack of these lines do not depend on the viewing angle.

Profiles of helium lines are of interest because the ejecta are rich in helium with 0.076~$\Msun$ (see \figref{fig:abundance}).
However, our spectrum synthesis simulation does not allow for a detailed investigation of this issue, because of the local thermal equilibrium assumption in the radiative transfer simulation \citep{1987ApJ...317..355H,1991ApJ...383..308L,2012MNRAS.422...70H}.
Therefore, we will discuss them in our future work, which is also interesting in relation to the similarity to SNe Iax, because a few SNe Iax are reported to show the He lines \citep{2013ApJ...767...57F}.

The right panel of \figref{fig:spectra} shows the dependence of the Ca II IR triplet profile on the viewing angle.
The spectra are redshifted or blueshifted, depending on the viewing angle, with velocities up to $\simeq12,000\,\kmpersec$, or with red/blueshift of $\Delta z \simeq \pm 0.054$.
This spectral shift is also predicted in the CO WD TDE model of \citet{2016ApJ...819....3M} with its velocity up to $\simeq\pm10,000\,\kmpersec$.
This is a distinguishable signature of the WD TDEs from the other transients such as SNe Ia.

% Todo: add discussion on viewing angle dependence on (dis)appearance
% Orbital energy of the debris of the WD spread due to the tidal disruption by the BH, which makes velocity dispersion of the debris,
% Orbital energy of the debris of the WD spread due to the tidal disruption by the BH, which makes velocity dispersion of the debris,
% In the tidal disruption by the BH,
The spectral shift is due to the intrinsic velocity of the bulk motion of the ejecta.
An orbital energy distribution of the tidal debris of the WD is spread due to the tidal disruption by the BH.
The unbound ejecta obtain positive orbital energy in this process, corresponding to the velocity of
% \begin{eqnarray}
%     \Delta \epsilon_t&\sim&\beta^n \frac{G \Mbh \Rwd} {R_t^{2}}\\
%     &\simeq& 3.1\times10^{17} \, \mathrm{erg}/\mathrm{g} \nonumber \\
%     &\;&\times\beta^n
% \left( \frac{\Rwd}{10^9\,\mathrm{cm}}\right)^{-1}
% \left( \frac{\Mbh}{10^{2.5}\,\Msun}\right)^{1/3}
% \left( \frac{\Mwd}{0.2\,\Msun}\right)^{2/3},
% \label{eq:spread_ene}
% \end{eqnarray}
% or the velocity dispersion of
\begin{eqnarray}
    % \Delta v_t &\sim& (2 \Delta \epsilon_t)^{1/2}\\
    v_t &\sim& \left(2 \beta^n \frac{G \Mbh \Rwd} {R_t^{2}} \right)^{1/2}\\
    &\simeq& 7.9 \times 10^3 \,\mathrm{km}\,\mathrm{s}^{-1} \nonumber \\
    &\;&\times \beta^{n/2}
\left (\frac{\Rwd}{10^9\,\mathrm{cm}}\right )^{-1/2}
    \left (\frac{\Mbh}{10^{2.5}\,\Msun}\right )^{1/6}
\left (\frac{\Mwd}{0.2\,\Msun}\right )^{1/3},
\label{eq:spread_vel}
\end{eqnarray}
where $n=2$ for a canonical model, while recent studies show that the value should be revised as $n=0$, which is the so-called "frozen-in approximation" \citep{2013MNRAS.435.1809S,2019MNRAS.485L.146S}.
The velocity is a little smaller than the velocity of the bulk motion, $\simeq 12,000\,\kmpersec$.
% We note that the release of the nuclear energy also enhances the spread of the orbital energy distributions.
We note that the release of the nuclear energy also increases the orbital energy of the ejecta.
The contributions by the tidal disruption and release of the nuclear energy result in the red/blueshifted spectra.
% We can see match between the velocity of the Doppler shifts and the velocity dispersion in the WD TDE.

Not only does the global spectral shift depend on the viewing angle, but absorption velocities do as well.
The right panel of \figref{fig:spectra} also shows the absorption velocities of the Ca II IR triplet line at $t\simeq16$~days for different viewing angle in the observer frame.
In the ejecta-comoving frame, the corresponding absorption velocity appears relatively smaller ($-11,000\,\kmpersec$) from the brightest angle ($\phi\simeq0.8$) than from the faintest angle ($\phi\simeq0.2$), where it is $-15,000\,\kmpersec$.
This is because the radial velocity with respect to the ejecta COM is smaller for the brightest angle than the faintest angle (see \figref{fig:distribution}).
This is qualitatively consistent with the behavior in the CO WD TDE \citet{2016ApJ...819....3M}.

\section{Discussions}
\label{sec:discussion}

\subsection{Fallback Components}
%%%%%%%%%%%%%%%%%%%%%%
%%fallback component%%
%%%%%%%%%%%%%%%%%%%%%%
In this study, we focus on emission from the unbound ejecta.
There could be an additional power source: the bound debris falling back onto the BH.
Here, we discuss which emission dominates over the other.
We can naively estimate the fallback luminosity from the mass fallback rate as
\begin{eqnarray}
    L_{\mathrm{fb}} &\sim& \eta \dot{M}_{\mathrm{fb}} c^2 \\
    % &\sim& \eta \dot{M}_{\mathrm{fb, p}} c^2 \left( \frac{t}{t_{\mathrm{fb}}} \right)^{-5/3}\\
    % &\simeq& 3 \times 10^{49}\, \mathrm{erg}/\mathrm{s} \,
    &\sim& 2 \times 10^{49}\, \mathrm{erg}\,\mathrm{s}^{-1} \,
    \left( \frac{t}{10^2\,\mathrm{s}} \right)^{-5/3} \nonumber \\
    &\;& \times
    \left( \frac{\eta}{0.1} \right)
    \left( \frac{\dot{M}_{\mathrm{fb,p}}}{10^{-4} \,\Msun\,\mathrm{s}^{-1}} \right)
    \left( \frac{t_{\mathrm{fb}}}{10^2\,\mathrm{s}} \right)^{5/3},
    \label{eq:Lfb}
\end{eqnarray}
where $\eta$ is the conversion efficiency.
We estimate the peak mass fallback rate $\dot{M}_{\mathrm{fb, p}}$ and fallback timescale $t_{\mathrm{fb}}$ from the results of the hydrodynamic simulation.
This luminosity is much larger than the luminosity of thermonuclear emission $L_{\mathrm{therm}}\simeq10^{42}\,\ergpersec$ around its peak $t\simeq5$~days.
However, the thermonuclear emission can be dominant if the fallback emission is limited by the Eddington luminosity,
% $L_{\mathrm{Edd}} \simeq 8 \times 10^{40} \,\mathrm{erg}\,\mathrm{s}^{-1} \left( \kappa / (0.2\,\mathrm{cm}^2\,\mathrm{g}^{-1} )\right) ^{-1} \left( \Mbh/10^{2.5} \,\Msun) \right)$
\begin{eqnarray}
    L_{\mathrm{Edd}} &\simeq& 8 \times 10^{40} \,\mathrm{erg}\,\mathrm{s}^{-1} \, \nonumber\\
    &\;& \times
    \left( \frac{\kappa}{0.2\,\mathrm{cm}^2\,\mathrm{g}^{-1}} \right) ^{-1}
    \left( \frac{\Mbh}{10^{2.5} \,\Msun} \right),
    \label{eq:LEdd}
\end{eqnarray}
where we take fiducial value of the opacity $\kappa$ as 0.2$\,\mathrm{cm}^2\,\mathrm{g}^{-1}$.
% We see $L_{\mathrm{Edd}} \ll L_{\mathrm{therm}} \ll L_{\mathrm{fb}}$ at $t\sim 5$~days, where $L_{\mathrm{therm}}$ is the bolometric luminosity of the emission from the unbound ejecta.
% Therefore, the thermonuclear emission dominate at the timescale of our interest unless super-Eddington emission from the bound debris are realized.
It is expected that super-Eddington emission from the fallback debris is observed only when relativistic jets are formed and are viewed on-axis (e.g. \citealt{2016ApJ...819....3M}).
Additionally, if the fallback debris forms an accretion disk and it emits thermal emission, its temperature would be $\sim10^6$~K \citep{2016ApJ...819....3M}.
In the optical wavelength, the disk thermal emission would be much smaller than the thermonuclear emission.
Note that although the fallback emission is not a main focus of this study, it is still an interesting target (e.g. \citealt{2017ApJ...841..132L}) and would be detectable with the eROSITA telescope \citep{2019MNRAS.489.5413M}.

\subsection{Possible Observational counterparts}
\label{sec:obs}
The synthetic spectra of the helium WD TDE have striking similarities to the spectra of SNe Iax (or 2002cx-like SNe Ia).
For example, the left panel of \figref{fig:spectra} shows that our day 12 spectrum (in the declining phase in the $V$-band) is very similar to the spectrum of SN Iax 2012Z around the $V$-band maximum \citep{2015A&A...573A...2S,2015ApJ...806..191Y}.
Both the spectra are dominated by Fe II, Fe III, Ca II, and lines from Fe-peak elements \citep{2004PASP..116..903B,2014A&A...561A.146S}.
This may indeed not be surprising; the helium WD TDE produces a mixture of Fe, Fe-peaks and IMEs, qualitatively similar to the weak/failed SN Ia model that explains the spectra of SNe Iax \citep{2013MNRAS.429.2287K}, except that the helium WD TDE has the higher expansion velocity.
The main differences are the nearly complete lack of Si II and S II in the TDE spectra, and the broader, more blue-shifted lines in the TDE.
% Adding the difference in the phases to match the spectral characteristics, it does not readily mean that the helium WD TDE could explain the properties of SNe Iax in general.
Another difference is seen in their photometric properties, especially in their declining timescales:
the helium WD TDE shows a more rapid decline ($\Delta m_{15} (\mathrm{B})\simeq4$~mag) than SNe Iax ($\Delta m_{15} (\mathrm{B})\simeq0.5\text{--}2.5$~mag).
However, given the diversity in the properties of SNe Iax \citep{2015A&A...573A...2S,2017hsn..book..375J}, it is an interesting possibility that a sub-population of SNe Iax may be explained by the helium WD TDE.

The helium WD TDE light curves show similar timescales and luminosities to the calcium-rich transients \citep{2010Natur.465..322P,2012ApJ...755..161K,2014MNRAS.437.1519V,2017ApJ...836...60L,2017ApJ...846...50M,2018ApJ...866...72D}, .Ia explosion candidates \citep{2010ApJ...723L..98K}, and other rapid transients \citep{2013ApJ...774...58D,2014ApJ...794...23D,2018Sci...362..201D,2018MNRAS.481..894P}.
We compare our model with observed transients of these classes, on their light curves (see \figref{fig:obs}) and spectra.

We do not find good matches between our model and the calcium-rich transients, .Ia explosion candidates, or rapid transients in \citet{2013ApJ...774...58D,2014ApJ...794...23D}, and \citet{2018Sci...362..201D}.
Compared with the calcium-rich transients, our model shows brighter and bluer emission around the peak.
This is because the $\nickel$ mass in our model, 0.030~$\Msun$ is larger than the calcium-rich transients, $\lesssim 0.015\,\Msun$.
Also, our model shows very weak silicon features, while a part of calcium-rich transients, such as PTF10iuv/SN2010et \citep{2012ApJ...755..161K} and SN2016hgs \citep{2018ApJ...866...72D}, show strong silicon features.
SN2010X, a .Ia explosion candidate, also shows strong silicon features and thus does not match to our model, while the light curves show good matches except for the $r$-band maximum: our models show brighter $r$-band maximum than SN2010X by $\simeq-$0.4~mag.
Our model shows fainter peak luminosity than any rapid transients reported in \citet{2013ApJ...774...58D,2014ApJ...794...23D} and \citet{2018Sci...362..201D}.

Our model can well explain multi-band light curves of two rapid transients among the samples presented by \citet{2018MNRAS.481..894P}.
% \figref{fig:obs} shows the good matches between our model light curves and those of observed transients.
% the comparison between light curves of our models and those of the two transients.
Physical offsets of these transients from the centers of their host galaxies are 3.26~kpc and 10.2~kpc for DES14S2plb and DES16S1dxu, respectively.
DES16S1dxu is noticeably distant from its host center, which may imply that an old stellar population is the source.
This may support its possible origin as a helium WD TDE, of which a star cluster containing an IMBH and WD is the plausible environment.
Although comparisons of spectra between our model and these transients are important to identify their origin(s) with certainty, such observational data are lacking.
Interestingly, they are two of the faintest transients among the rapid transients presented by \citet{2013ApJ...774...58D,2014ApJ...794...23D}, \citet{2018Sci...362..201D}, and \citet{2018MNRAS.481..894P}.
\citet{2018MNRAS.481..894P} reported a much larger sample of rapid transients than the other studies so that they would be able to find the faintest transients with relatively low event rates.
% \citet{2018MNRAS.481..894P} reported a larger sample of rapid transients with the redshifts of their host galaxies, 37 transients, than \citet{2014ApJ...794...23D}, 10 transients.
% This may be the reason why our model does not match the transients in \citet{2014ApJ...794...23D}: faint transients are missed with this limited sample.
To find helium WD TDEs more certainly, we encourage searching for a larger number of faint and rapid transients and performing rapid spectroscopic follow-ups.

%%%%%%%%%%%%%%%%%%%%%%
%%compare with obs%%%%
%%%%%%%%%%%%%%%%%%%%%%
% DLR

\subsection{Variety}
\label{sec:variety}
%%%%%%%%%%%%%%%%%%%%%%
%%%%%% variety %%%%%%%
%%%%%%%%%%%%%%%%%%%%%%

We consider a single parameter set of the WD TDE in this study, namely $\Mbh = 10^{2.5}\,\Msun$, $\Mwd=0.2\,\Msun$, and $\beta=5$.
\citet{2018MNRAS.477.3449K} showed that nucleosynthesis and dynamics of ejecta, such as the $\nickel$ mass and ejecta mass, in WD TDEs have a large variety depending on these parameters.
Therefore, we intuitively expect a large variety of luminosity, timescale, and other features of the thermonuclear emission from WD TDEs for other parameter cases.
Although our particular model in this study does not explain observed SNe Iax, calcium-rich transients, .Ia explosion candidates, or other rapid transients, they may be explained using other parameter cases.
For example, we can expect even more rapidly evolving transients than the model in this study if we consider less-massive WDs, because it would lead to less-massive ejecta and smaller diffusion timescale.
Such a model may be able to explain a very rapidly declining type I transient SN2019bkc/ATLAS19dqr (\citealt{2019arXiv190905567P,2020ApJ...889L...6C}).
If we consider more-massive WDs, instead, more slowly evolving transients would appear and may explain a sub-population of SNe Iax.
In the future, we plan to study the thermonuclear emission from WD TDEs to establish a range of the parameter space.

\acknowledgements

The authors thank Kazumi Kashiyama, Toshikazu Shigeyama, Masaomi Tanaka, and Takashi Moriya for fruitful discussions.
We also thank Morgan MacLeod and Miika Pursiainen for kindly providing us with their data.
K.K. is supported by the Advanced Leading Graduate Course for Photon Science.
This work is supported by JSPS KAKENHI Grant Number 17H02864, 17H06360, 18H04585, 18H05223, 18J20547, 19K03907.
Numerical calculations in this work were carried out on Cray XC50/XC30 at Center for Computational Astrophysics, National Astronomical Observatory of Japan and on Cray XC40 at the Yukawa Institute Computer Facility.

\bibliography{ref}{}
\bibliographystyle{aasjournal}

\end{document}